# Studies of Nanotube Channeling for Efficient Beam Scraping at Accelerators

V.M. Biryukov[1,•], S. Bellucci[2]

1 *Institute for High Energy Physics, Pobedy 1, Protvino, RU-142281, Russia*
2 *INFN - Laboratori Nazionali di Frascati, P.O. Box 13, 00044 Frascati, Italy*

**Abstract**

While particle beam steering (and in particular, "scraping") in accelerators by bent channeling crystals is an established technique extensively tested at IHEP Protvino and other major high-energy labs, an interesting question is how one could improve channeling capabilities by applying modern nanotechnology. Theoretical research of nanotube channeling was in progress over recent years. In this work, we assess potential benefits from nanotube channeling for real accelerator systems. We report simulation studies of channeling in nanostructured material (carbon SWNT and MWNT) tested for possible serving as a primary scraper for the collimation systems of hadron colliders. The advantages of nanostructured material as a potential choice for a primary scraper in a high-energy accelerator such as LHC or the Tevatron are discussed in comparison to crystal lattices and amorphous material. We evaluate physical processes relevant to this application and reveal nanotechnology requirements.

## 1. Introduction

The technique of particle beam channeling by bent crystals is well established at accelerators [1] and demonstrated in the energy range spanning over six decades [1-14]. For instance, bent crystals are largely used for extraction of 70-GeV protons at IHEP (Protvino) [10-12,15,16] with efficiency reaching 85% at intensity of $10^{12}$ particle per second, steered by silicon crystal of just 2 mm in length [11]. Following the successful experiments on crystal channeling in accelerator rings, there has been a strong interest to apply channeling technique in a TeV-range collider for beam extraction or collimation [2-4,7-9,17-19].

While bent crystals have found a good niche for their application, there is a permanent interest in upgrading the capabilities of channeling structures, in order to increase the efficiency of the technique and to widen its span of applications. There has been large interest recently [20-24] to the possibility of channeling in the materials produced by nanotechnology.

---

[•] Corresponding author. Web: http://www1.ihep.su/~biryukov/. E-mail: biryukov@mx.ihep.su

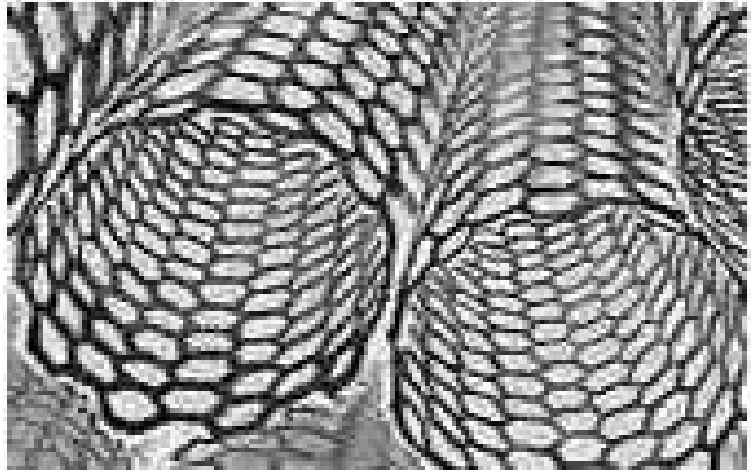

**Fig. 1. Schematic view of nanostructured lattice made of SWNT**

Carbon nanotubes are cylindrical molecules with a typical diameter of order 1 nm for single-wall nanotubes (SWNT) or a few tens nm for multi-wall nanotubes (MWNT) and a length of many microns [25]. They are made of carbon atoms and can be thought of as a graphene sheet rolled around a cylinder, Fig.1. Creating efficient channeling structures - from single crystals to nanotubes - might have a significant effect onto the accelerator world. With nanotechnology in hands, there is a principle possibility to engineer a channeling lattice from wanted atoms and of wanted configuration, to some extent. Therefore, it would be interesting to understand what kind of channeling structure we would prefer if we could choose. Ideally, we would like to trap particles in two dimensions; to have the channel walls made of densely packed atoms; to make channel size of our choice; to build it with atoms of our choice. This is indeed close to what we get from nanotechnology nowadays.

## 2. Halo scraping in colliders

Here we assess potential benefits from nanotube channeling for real accelerator systems on the example of halo scraping in colliders. Classic two-stage collimation system for loss localisation in accelerators typically uses a small scattering target as a primary element and a bulk absorber as secondary element [26]. Normally in colliders and storage rings the halo diffusion is rather slow, therefore the first touch of a halo particle with the aperture-restricting collimator is rather a glancing touch, with impact parameters of the order of a micron or less. Unfortunately, particles interacting near to the surface of a solid block can be scattered back in the beam pipe; hence they can be lost far from the cleaning area. The role of the primary element is to give a substantial angular kick to the incoming particles in order to increase the impact parameter on the secondary element, which is generally placed in the optimum position to intercept transverse or longitudinal beam halo.

An amorphous primary target scatters the impinging particles in all possible directions. Ideally, one would prefer to use a "smart target" which kicks all particles in only one direction: for instance, only in radial plane, only outward, and only into the preferred angular range corresponding to the center of the absorber (to exclude escapes).

Fig.2 shows the angular distribution of 7 TeV protons downstream of an amorphous carbon target due to elastic scattering on carbon nuclei. In accelerator, this Gaussian angular distribution transforms into a broad radial distribution to be partially intercepted by absorber of the scraping system. Obviously, the distribution would be peaked at the edge of the intercepting material. The edge leak of outscattered particles is a major source of inefficiency in this kind of system. However, the LHC scraping system must have efficiency of interception on the order of 0.9999!

Bent crystal was studied [11,13,18,19] as the first idea for a smart target: it traps particles and forwards them into the desired direction. Here, the scattering process on single atoms of an amorphous target becomes the selective and coherent scattering on atomic planes of an aligned mono-crystal.

Fig. 2 shows also an example of the angular distribution of 7 TeV protons channeled and bent 10 μrad by a carbon nanostructure (made of SWNT with 1.1 nm diameter) simulated in the approach presented and discussed in the next section. In this case, all channeled protons can be safely intercepted downstream and nonchanneled protons scatter in the way similar to amorphous target. If, e.g., 90% of halo protons were channeled and just 10% scattered incoherently in the primary target, then the inefficiency (the edge leak) of the scraping system would reduce by a factor of 10. Below we study in simulations what efficiency can be achieved here with channeling nanostructures.

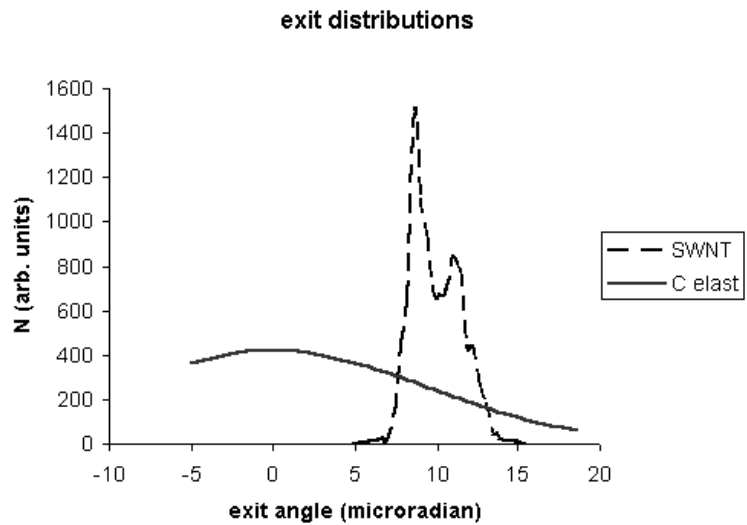

**Fig 2. Angular distributions downstream of a primary target due to elastic nuclear scattering on the atoms of amorphous carbon target and due to channeling in carbon SWNT target.**

**3. Nanostructured primary scraper**

For simulation studies of channeling in nanostructured material (carbon SWNT and MWNT) tested for possible serving as a primary scraper for the collimation systems of hadron colliders we started from the same computer model [27,28] as been used for simulation of a crystal scraping in the Large Hadron Collider and replaced bent crystal by a nanostructured primary scraper bent a small angle to steer the impinging halo particles. Simulations were done in the LHC both at the collision energy of 7 TeV and at the injection energy of 450 GeV. In the model, a nanostructured scraper was positioned as a primary element at a horizontal coordinate of $6\sigma$ in the halo of the LHC beam, on one of the locations presently chosen for amorphous primary elements of the LHC collimation system design [29-31]. The LHC lattice functions were taken corresponding to this position: $\alpha_x$=1.782 and $\beta_x$=119 m in the horizontal plane, and $\alpha_y$ = −2.02 and $\beta_y$ =143.6 in the vertical plane. The $\alpha$ and $\beta$ functions define the transfer matrix which transforms the angles and coordinates of the particle revolving in the accelerator ring before the particle can interact again with the scraper; further details can be found e.g. in ref. [1] or any accelerator textbook.

The particles were tracked through the curved lattice of nanotubes in the approach with a continuous potential described in more detail elsewhere [32]. We varied nanotube parameters such as the length, diameter, bending, alignment angle, and the structure: either SWNT or MWNT. We observed the efficiency of channeling, i.e., the number of the particles deflected at the full bending angle of the lattice, taking into account many turns in the LHC ring and multiple encounters with the nanostructured scraper.

Only the halo particles incident into the opening of any particular nanotube within its angular acceptance could have chances to be channeled and bent some angle. Particles falling between the nanotubes in the lattice were not channeled. The nonchanneled particles still experience coherent and incoherent scatterings in the bulk of nanostructured scraper material, including a chance of inelastic nuclear interaction. For these processes we assumed the density of nanostructured target equal to that of graphite. We assumed that we have no amorphous inclusions in the target, so the incident particles can be channeled already on the very first encounter. The inner diameter of simulated MWNT was 2 nm (few nm is a typical value for MWNT).

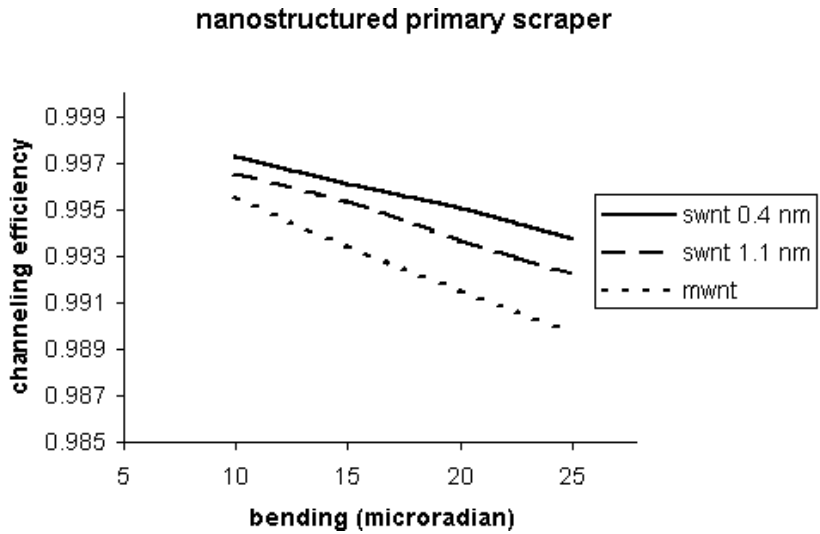

**Fig. 3. The overall (multi-pass) channeling efficiency *F* of nano-structured scraper shown as a function of bending angle for SWNT (0.4 and 1.1 nm diameter) and MWNT at 7 TeV.**

Fig. 3 shows the computed channeling efficiency as a function of the bending angle of the nanostructured lattice for 7 TeV protons. For each bending angle the optimal length of the structure was found. The optimal length is about 1 mm for 10-µrad bending. High efficiency of channeling can be obtained with the same structure both at 7 TeV and at 450 GeV (injection energy in the LHC ring). The efficiency *F* can be as high as 0.99-0.997 for the small bendings studied in Fig. 3, thanks to the low divergence of halo particles incident onto aligned structure and to a large extent thanks to the possibility of multiple encounters of circulating halo particles with channeling structure. Fig. 4 shows the same data plotted as a "background reduction factor" *1/(1-F)*. We expect the intensity of the halo (causing the background irradiation in the machine) to be reduced by this factor thanks to the effect of channeling. If all channeled particles were fully intercepted by the secondary collimator, then only non-channeled particles could contribute to the background in the accelerator; therefore the scraping efficiency improves by the above-mentioned factor.

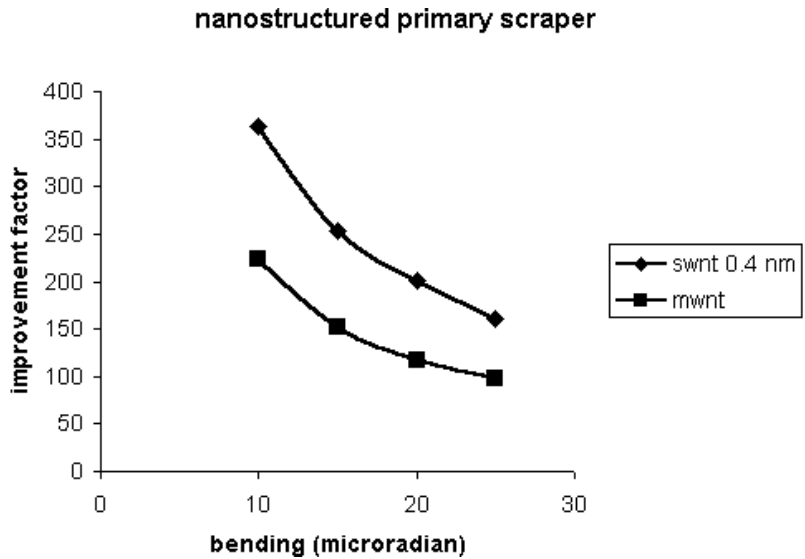

**Fig. 4. The same data as in previous Figure, plotted as an LHC background reduction factor 1/(1-*F*) as a function of bending angle, for carbon SWNT and MWNT.**

For an efficient operation, the nanostructure must be oriented parallel to the envelope of the circulating beam. According to simulations, for best channeling efficiency, the nanostructure has to be aligned with the accuracy of about ±3 µrad. Obviously, this also means that the nanotubes within the lattice must be parallel to each other with the same precision. Notice that conventional beam collimation elements at the LHC have to be aligned with 15 μrad accuracy [29-31].

An important issue is the power deposit through particle hits inducing thermal shock, radiation damage and eventually reducing the structure lifetime. The halo intensity in LHC is order of $10^9$ proton/s. In typical channeling tests at IHEP U-70, crystal channeled $\sim 10^{12}$ protons in a spill of 0.5-1 s duration [11]. One of the IHEP crystals served in the vacuum chamber of U-70 over 10 years, from 1989 to 1999, delivering beam to particle physicists, until a new crystal replaced it. For nanostructures there is not enough data, although damage processes are being evaluated [33]. However crystals show a huge margin of safety in lattice lifetime and thus raise hope for similar stability of nanostructures.

## 4. Conclusions

From physics standpoint, channeling nano-structures could be an interesting and very efficient choice for the primary element of the scraping system in a collider or storage ring. Their use could suppress the accelerator-related background by a factor of 100-300 if all technological issues are resolved. To be efficient at the energy as high as 7 TeV of the LHC, the nanostructures should be as long as about 1 mm; their thickness on the order of 1 μm is sufficient for effective channeling of halo particles. Another major requirement is alignment as good as about ±3 μrad at 7 TeV. At lower energy the requirements are much less stringent; e.g., at 70 GeV the nanostructures as short as some 20 μm produce significant steering effect on particles [34].

**Acknowledgements**.

This work was partially supported by INFN - Gruppo V, as NANO experiment, and by INTAS-CERN grants 132-2000 and 03-52-6155.